\documentclass[twocolumn]{aastex63}
\usepackage{natbib,amsmath,scalerel,placeins}

\usepackage{graphicx}
\usepackage{txfonts}
\usepackage{xspace}
\usepackage{color}
\usepackage{url}
\usepackage{hyperref}
\usepackage{comment}
\usepackage[flushleft]{threeparttable}

\def\arcdeg{\hbox{$^\circ$}}

\def\arcsec{\hbox{$^{\prime\prime}$}}
\def\flux{erg\,s$^{-1}$\,cm$^{-2}$}

\def\lum{erg s$^{-1}$}

\def\nustar{\textit{NuSTAR}\xspace}
\def\swiftxrt{\textit{Swift}/XRT\xspace}

\def\cgro{\textit{CGRO}\xspace}

\def\srg{\textit{SRG}\xspace}
\def\art{\textit{SRG}/ART-XC\xspace}

\def\gro{GRO\,J1008$-$57\xspace}

\newcommand {\be}{\begin {equation}}
\newcommand {\ee}{\end {equation}}

\def\a{$^{\mbox{\small a}}$}
\def\b{$^{\mbox{\small b}}$}
\def\c{$^{\mbox{\small c}}$}
\def\d{$^{\mbox{\small d}}$}
\def\e{$^{\mbox{\small e}}$}

\received{XXX, 2020}
\revised{XXX, 2020}
\accepted{XXX}
\submitjournal{ApJ}

\shorttitle{Low-level accretion in \gro}
\shortauthors{Lutovinov et al.}

\begin{document}

\title{{\it SRG}/ART-XC and {\it NuSTAR} observations of the X-ray pulsar \gro in the lowest luminosity state}

\correspondingauthor{A. Lutovinov}
\email{aal@iki.rssi.ru}

\author{A. Lutovinov}
\affiliation{Space Research Institute, Russian Academy of Sciences, Profsoyuznaya 84/32, 117997 Moscow, Russia}

\author{S. Tsygankov}
\affiliation{Department of Physics and Astronomy, FI-20014 University of Turku, Finland}
\affiliation{Space Research Institute, Russian Academy of Sciences, Profsoyuznaya 84/32, 117997 Moscow, Russia}

\author{S. Molkov}
\affiliation{Space Research Institute, Russian Academy of Sciences, Profsoyuznaya 84/32, 117997 Moscow, Russia}

\author{V. Doroshenko}
\affiliation{Institute for Astronomy and Astrophysics, University of
T\"ubingen, Sand 1, 72026 T\"ubingen, Germany}
\affiliation{Space Research Institute, Russian Academy of Sciences, Profsoyuznaya 84/32, 117997 Moscow, Russia}

\author{A. Mushtukov}
\affiliation{Leiden Observatory, Leiden University, NL-2300RA Leiden, The Netherlands}
\affiliation{Space Research Institute, Russian Academy of Sciences, Profsoyuznaya 84/32, 117997 Moscow, Russia}
\affiliation{Pulkovo Observatory, Russian Academy of Sciences, Saint Petersburg 196140, Russia}

\author{V. Arefiev}
\affiliation{Space Research Institute, Russian Academy of Sciences, Profsoyuznaya 84/32, 117997 Moscow, Russia}

\author{I. Lapshov}
\affiliation{Space Research Institute, Russian Academy of Sciences, Profsoyuznaya 84/32, 117997 Moscow, Russia}

\author{A. Tkachenko}
\affiliation{Space Research Institute, Russian Academy of Sciences, Profsoyuznaya 84/32, 117997 Moscow, Russia}

\author{M. Pavlinsky}
\affiliation{Space Research Institute, Russian Academy of Sciences, Profsoyuznaya 84/32, 117997 Moscow, Russia}


\begin{abstract}
We report results of the first broadband observation of the transient X-ray pulsar \gro performed in the quiescent state. Observations were conducted quasi-simultaneously with the {\it Mikhail Pavlinsky} ART-XC telescope on board {\it SRG} and {\it NuSTAR} right before the beginning of a Type I outburst. \gro was detected in the state with the lowest observed luminosity around several $\times 10^{34}$ \lum and consequently accreting from the cold disk. Timing analysis allowed to significantly detect pulsations during this state for the first time. The observed pulsed fraction of about 20\% is, however, almost three times lower than in brighter states when the accretion proceeds through the standard disk. We traced the evolution of the broadband spectrum of the source on a scale of three orders of magnitude in luminosity and found that at the lowest luminosities the spectrum transforms into the double-hump structure similarly to other X-ray pulsars  accreting at low luminosities (X Persei, GX 304--1, A~0535$+$262) reinforcing conclusion that this spectral shape is typical for these objects.
\end{abstract}

\keywords{pulsars: individual (\gro) -- stars: neutron -- X-rays: binaries}



\section{Introduction}
\label{sec:intro}
A significant progress in studies of strongly magnetized neutron stars in binary systems (a.k.a. X-ray pulsars, XRPs) has been made over the last years due to extraordinary technical capabilities of the currently operating X-ray observatories. As a result, the range of observable luminosities from such objects extended over nine orders of magnitude from $L_{\rm X}\sim10^{40}-10^{41}$ \lum\ for pulsating ultra luminous X-ray sources \citep{2014Natur.514..202B,2016ApJ...831L..14F,2017Sci...355..817I,2020ApJ...895...60R,2020MNRAS.491.1857D} to $L_{\rm X}\sim10^{32}-10^{33}$ \lum\ for XRPs in quiescent state \citep[see e.g.,][]{2002ApJ...580..389C,2014A&A...561A..96D,2016MNRAS.463L..46W,2017MNRAS.470..126T}, allowing to study absolutely different regimes of accretion onto highly magnetized neutron stars (NSs). In addition to the improved technical characteristics, the flexibility in arrangement for long-term monitoring campaigns played a major role. For instance, {\it Swift}/XRT data allowed to discover transitions of several XRPs to the propeller regime, caused by the centrifugal inhibition of accretion \citep[see, e.g.,][]{2016A&A...593A..16T,2017ApJ...834..209L,2019MNRAS.485..770L,2019MNRAS.490.3355S}. For slower pulsars a new regime of the accretion, from a low-ionized (cold) accretion disk, was discovered \citep{2017A&A...608A..17T}. This regime is characterized by the transition of accretion to a quasi-stable state with the luminosity of the order of $L_{\rm X}\sim10^{34}-10^{35}$ \lum. Broadband observations of several pulsars at this low luminosity level demonstrated also a rather complex shape consisting of two components -- a low-energy thermal component and a high-energy one, presumably associated with the cyclotron radiation \citep{2019MNRAS.483L.144T,2019MNRAS.487L..30T,2020arXiv200613596M}.

Since physics of interaction of the cold disk with the strong magnetic field of the NS is still unexplored, any additional observations in this state are of vital importance.
The first pulsar where the accretion from the cold disk was discovered is \gro, a long-period ($P\sim94$ s) transient XRP in the binary system with the Be star as a normal companion and with an orbital period of $\sim250$ days \citep{2006ATel..940....1L,2007MNRAS.378.1427C,2012ATel.4564....1K}. The pulsar was discovered in 1993 with the \cgro\ observatory \citep{1993IAUC.5836....1S}. The neutron star is strongly magnetized with magnetic field of $\sim8\times10^{12}$~G estimated from the cyclotron line discovered around 75--90 keV \citep{1999ApJ...512..920S,2013ATel.4759....1Y,2020ApJ...899L..19G}.

\gro shows both giant irregular type II outbursts and regular type I outbursts associated with the passage of a neutron star through the periastron. A typical duration of the type I outburst is about 30 days with the maximum luminosity of $L_{\rm X}\sim10^{37}$\,\lum\ (assuming distance to the source of $\sim5.8$ kpc, \citealt{2012A&A...539A.114R}) and decay time of several days. A transition to the regime of the accretion from the cold disk happens around one month after the peak of the outburst when the luminosity drops to $L_{\rm X}\sim10^{35}$\,\lum\ and then remains almost constant slowly decreasing until the next outburst \citep{2017A&A...608A..17T}.

The most puzzling property of this state in \gro is a significant decrease of the pulsed fraction, which dropped from $\sim50\%$ in the bright state to below $\sim20\%$ (upper limit) in the low one \citep{2017A&A...608A..17T}. Also the spectrum shape in the low state in a wider energy range (above 10~keV) is still unknown, since no sensitive observations in broad energy band were performed until now.
In this work we present results of the dedicated observations of \gro in the regime of the accretion from the cold disk done with the {\it Mikhail Pavlinsky} ART-XC telescope on board the \srg mission, and the \nustar observatory. High quality of the data allowed to detect for the first time the pulsations in such a low state and to trace an evolution of the source broad band spectrum in a wide range of luminosities.

\begin{table}
    \centering
    \caption{List of \gro observations utilized in this work.}
    \begin{tabular}{ccccc}
    \hline
    ObsID & Start date & Start MJD & Exp. (ks) & $P_{\rm spin}$, s $^a$\\
    \hline
    \multicolumn{5}{c}{\nustar}  \\
    80001001002 & 2012-11-30 & 56261.36 & 12.5 & 93.5727(2) \\
    90001003002 & 2014-12-03 & 56994.81 & 22.3 & 93.4429(4) \\
    90001003004 & 2015-01-27 & 57049.74 & 12.5 & 93.3606(2) \\
    90401338002 & 2018-10-17 & 58408.67 & 45.3 & 93.2207(2) \\
    90501357002 & 2020-01-01 & 58849.79 & 45.2 & 93.20(1) \\
    \multicolumn{5}{c}{\art}  \\
    -- & 2019-12-11 & 58828.60 & 30.8 & 93.24(3)  \\
    \hline
    \end{tabular}
\hspace{-1cm} $^a$ barycentric and binary corrections were applied. Orbital parameters were
taken from \citet{2013A&A...555A..95K}
    \label{tab:data}
\end{table}

\section{Observations and data analysis} \label{sec:data}

The \srg observatory was successively launched to the orbit with the Proton rocket on Jul 13, 2019. The scientific payload includes two X-ray telescopes, eRosita (working energy range of $0.2-10$ keV) and the {\it Mikhail Pavlinsky} ART-XC telescope (working energy range of $4-30$ keV), which are primary designed to carry out the all-sky survey with unprecedented sensitivity.

ART-XC is a grazing incidence focusing X-ray telescope which provides imaging, timing and spectroscopy in the 4-30 keV energy range \citep{ART2021}. The telescope consists of seven identical modules with the total effective area of $\sim450$\,cm$^2$ at 6 keV, angular resolution of $45$\arcsec, energy resolution of $1.4$ keV at 6 keV and timing resolution of 23 $\mu$s.

\begin{figure}
\centering
\vspace{-2.7cm}
\includegraphics[width=0.95\columnwidth,bb=20 170 530 690]{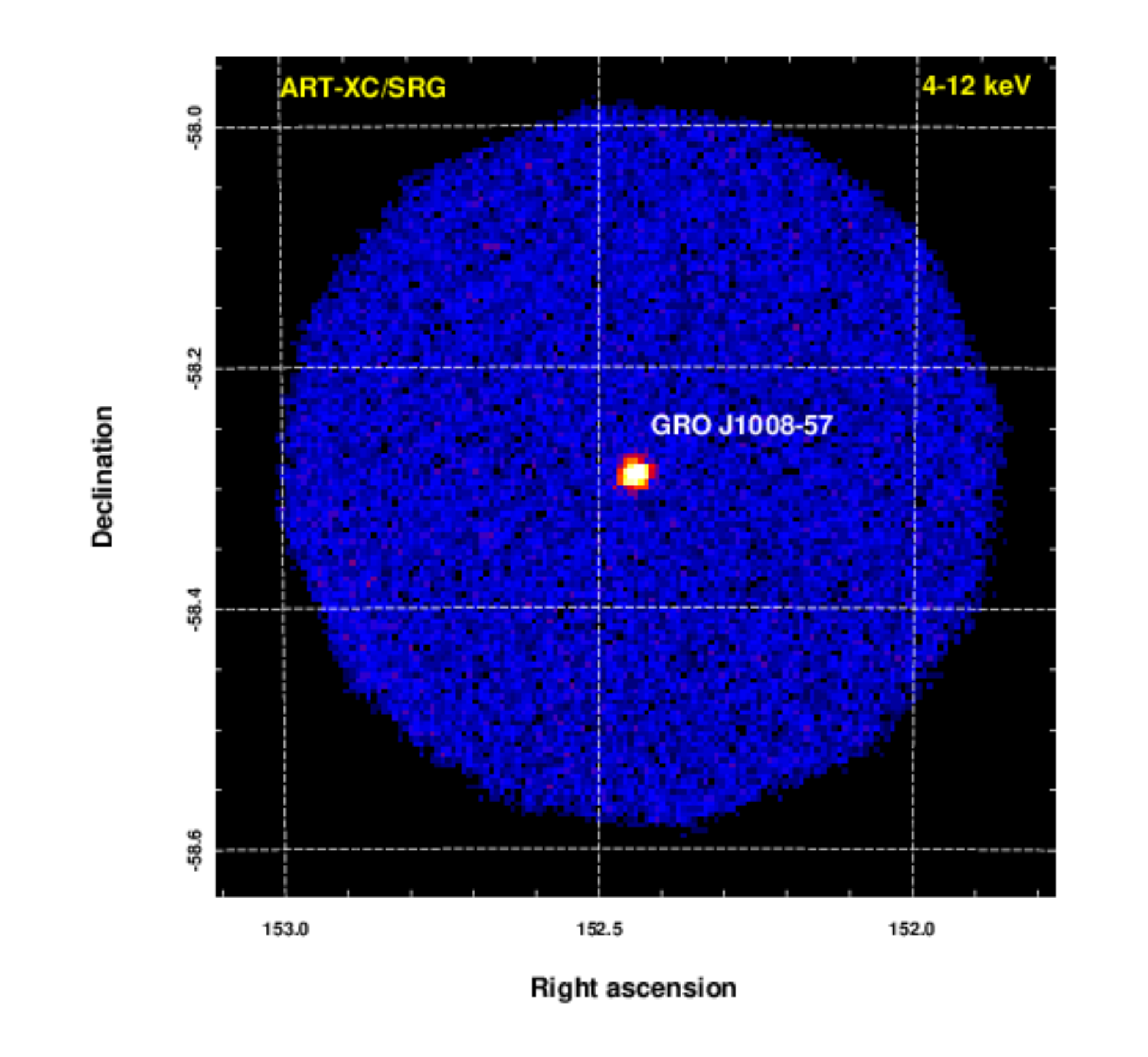}
\includegraphics[width=0.95\columnwidth,bb=0 60 260 260]{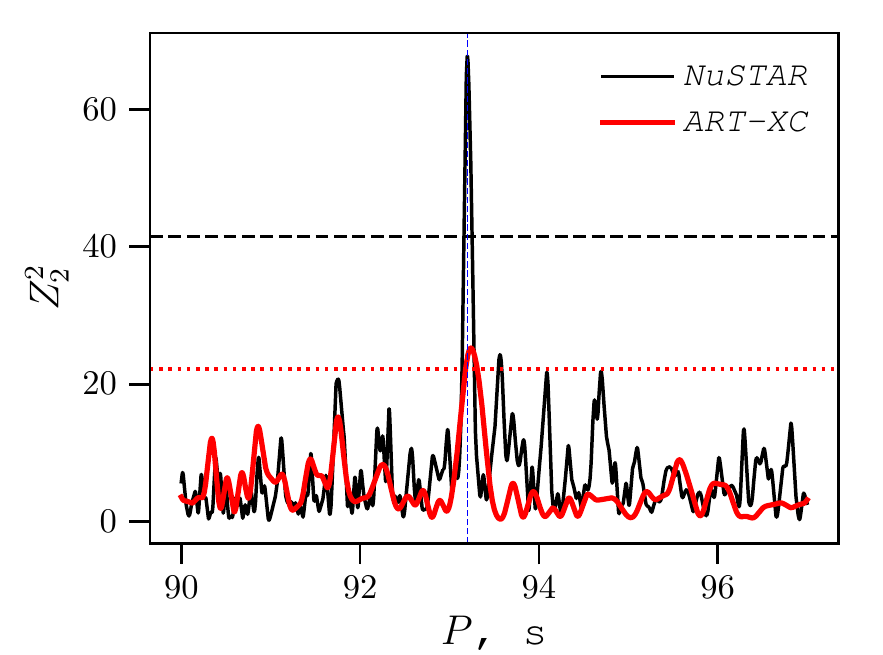}
\includegraphics[width=0.95\columnwidth,bb=0 40 530 420]{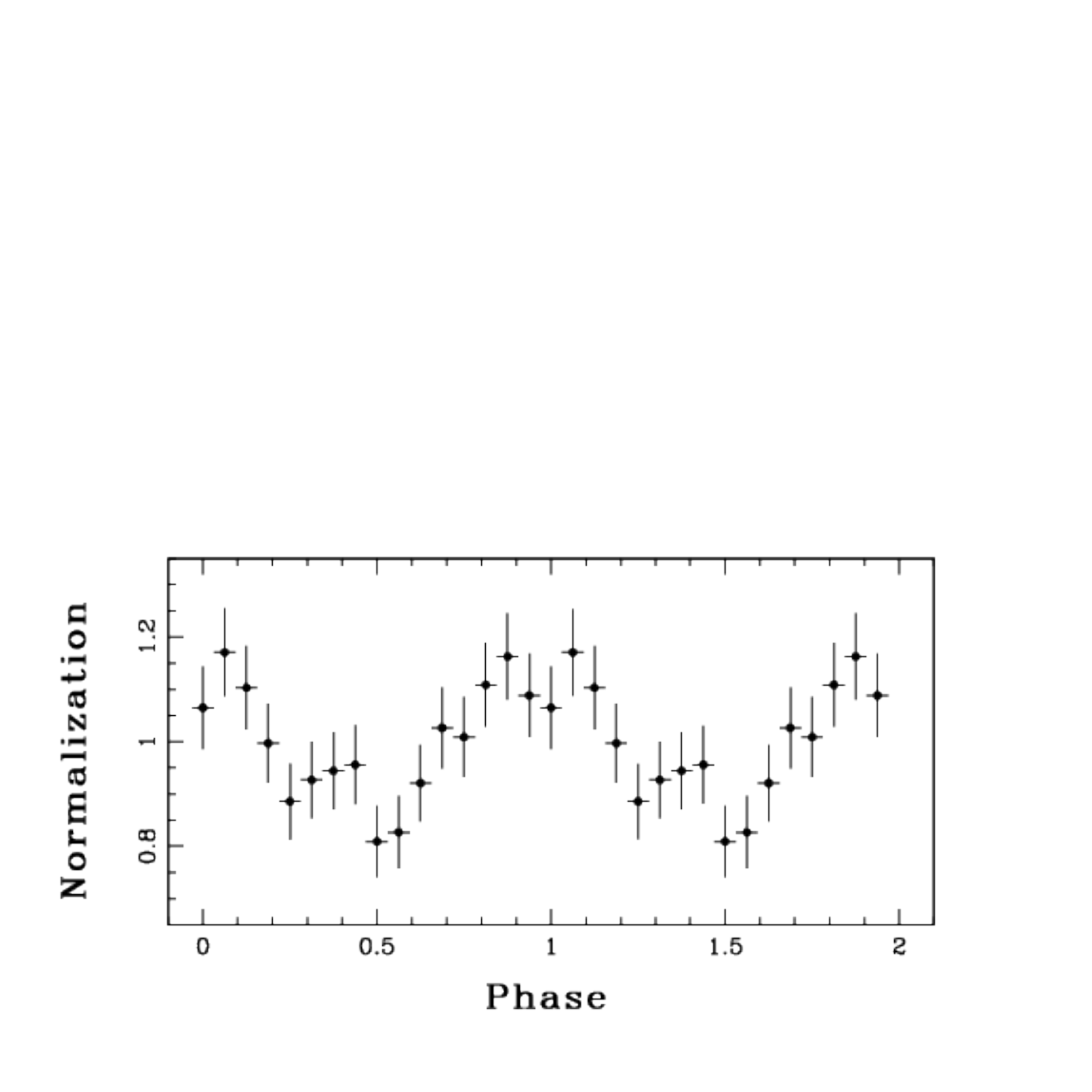}
\caption{{\it Upper panel} \art\ detection of \gro in the low luminosity state. {\it middle panel} $Z^2_2$ periodograms for the \art\ and \nustar\ observations are shown by red and black lines (together with the expected $3\sigma$ noise upper limit estimated as described in \cite{1994MNRAS.268..709B} assuming 22\% pulsed fraction). {\it Bottom panel} The source light curve measured with ART-XC in the 4-12 keV energy band and folded with the period of 93.24 s (middle panel).}
\label{fig:artima}
\end{figure}

During the calibration and performance verification phase (CalPV) conducted soon after the launch, the observatory performed a number of scanning and pointing observations of multiple fields and sources, including \gro. This source was observed on Dec 11, 2019 in the pointing mode with the total exposure of 30 ks. The ART-XC telescope was a primary instrument in this observation and worked in a nominal regime. ART-XC data were processed with the analysis software {\sc artproducts} v0.9 with the {\sc caldb} version 20200401. Brief description of the ART-XC analysis software can be found in \citet{ART2021}.

Preliminary analysis of ART-XC data showed that \gro\ was significantly detected (Fig.~\ref{fig:artima} upper panel) with the flux of $\sim3 \times 10^{-12}$ \flux\ in the $4-12$ keV energy band, that is, close to the minimal values registered from the source \citep{2017A&A...608A..17T}. Pulsations with the period of 93.24\,s from the source were confidently detected with ART-XC with $\sim3\sigma$ significance (see middle and bottom panels in Fig.~\ref{fig:artima}). We note that, as shown in the same figure, pulsations with the same period are also detected with \nustar, where significance of the pulsed signal detection is even higher with false alarm probability of $\sim2.3\times10^{-6}$.
It is the first time when pulsations have been detected from the source while in the ``cold disk'' state. To better characterize spectral and timing properties of the pulsar in the low luminosity state in the broad energy band, we also requested observations with the \nustar\ observatory. Thanks to the efforts of the \textit{NuSTAR} team, observations had been carried out just twenty days after the \art\ pointing. The preliminary analysis showed that between ART-XC and \textit{NuSTAR} observations the source flux was increased by $\sim30$\%. Nevertheless considering the slow evolution of flux and spectral parameters of the source quiescence \citep{2017A&A...608A..17T}, the \art\ and \nustar\ observations can be considered quasi-simultaneous.

\begin{figure}
\centering
\vspace{-2.0cm}
\includegraphics[width=0.97\columnwidth,bb=20 10 550 710]{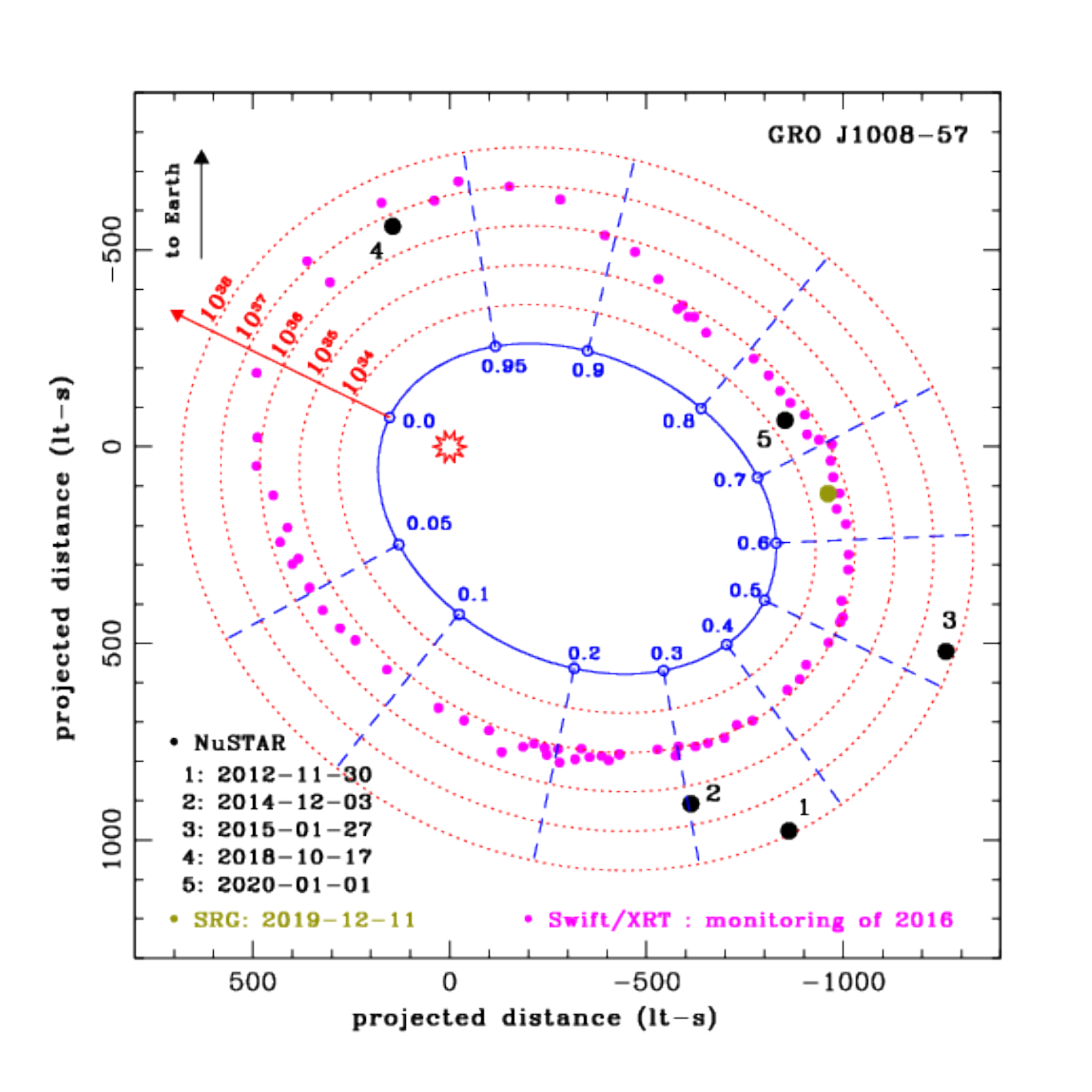}
\caption{Two-dimensional bolometrically corrected orbital light curve of \gro obtained in 2016 with {\it Swift}/XRT (magenta filled circles). Bolometric fluxes that were measured with \srg/ART-XC and \nustar observations are shown with light green and black colors points, respectively. Orbital parameters were taken from \citet{2013A&A...555A..95K}.
}
\label{fig:lc}
\end{figure}

The {\it NuSTAR} observatory consists of two identical X-ray telescope modules, referred to as FPMA and FPMB \citep{2013ApJ...770..103H}. It provides X-ray imaging, spectroscopy and timing capabilities in the energy range of 3-79~keV. {\it NuSTAR} performed an observation of \gro on Jan 1, 2020 (ObsID 90501357002) in the state with approximately the same intensity as during the \art\ observation.

To compare spectral and timing properties of \gro in the low state with those during the high state, we also re-analyzed archival \nustar observations of the source. In particular, observations performed on Nov 30 2012 (ObsID 80001001002), Dec 3 2014 (ObsID 90001003002), Jan 27 2015 (ObsID 90001003004), and Oct 17 2018 (ObsID 90401338002) were used. All {\it NuSTAR} data were processed with the standard {\it  NuSTAR} Data Analysis Software ({\sc nustardas}) v1.8.0 provided under {\sc heasoft} v6.25 with the {\sc caldb} version 20200826.

Finally, \swiftxrt\ observed the source simultaneously with \nustar on Jan 1, 2020 in the photon counting mode with the exposure of $\sim2$ ks (ObsID 00089019001). We used these data to extend the source spectrum into low energies and to compare directly with ART-XC results. Other \swiftxrt\ observations simultaneous with above mentioned \nustar ones were used to study the broad band spectra of \gro. The  data were processed using the online tools\footnote{\url{http://www.swift.ac.uk/user_objects/}} \citep[][]{2009MNRAS.397.1177E} provided by the UK Swift Science Data Centre.

All spectra obtained in four brightest states were grouped to have at least 25 counts per bin using the {\sc grppha} tool. Spectra obtained by \nustar, \swiftxrt\ and \art\ in the lowest state were binned to have at least one count per energy bin and modeled using W-statistic \citep{1979ApJ...230..274W}. The final data analysis (timing and spectral) was performed with the {\sc heasoft 6.25} software package. All uncertainties are quoted at the $1\sigma$ confidence level, if not stated otherwise. Finally, the spectra were also rebinned for plotting in {\sc Xspec}.

\bigskip
~~
\begin{figure}
\centering
\vspace{0.2cm}
\includegraphics[width=0.97\columnwidth,bb=35 170 550 700,clip]{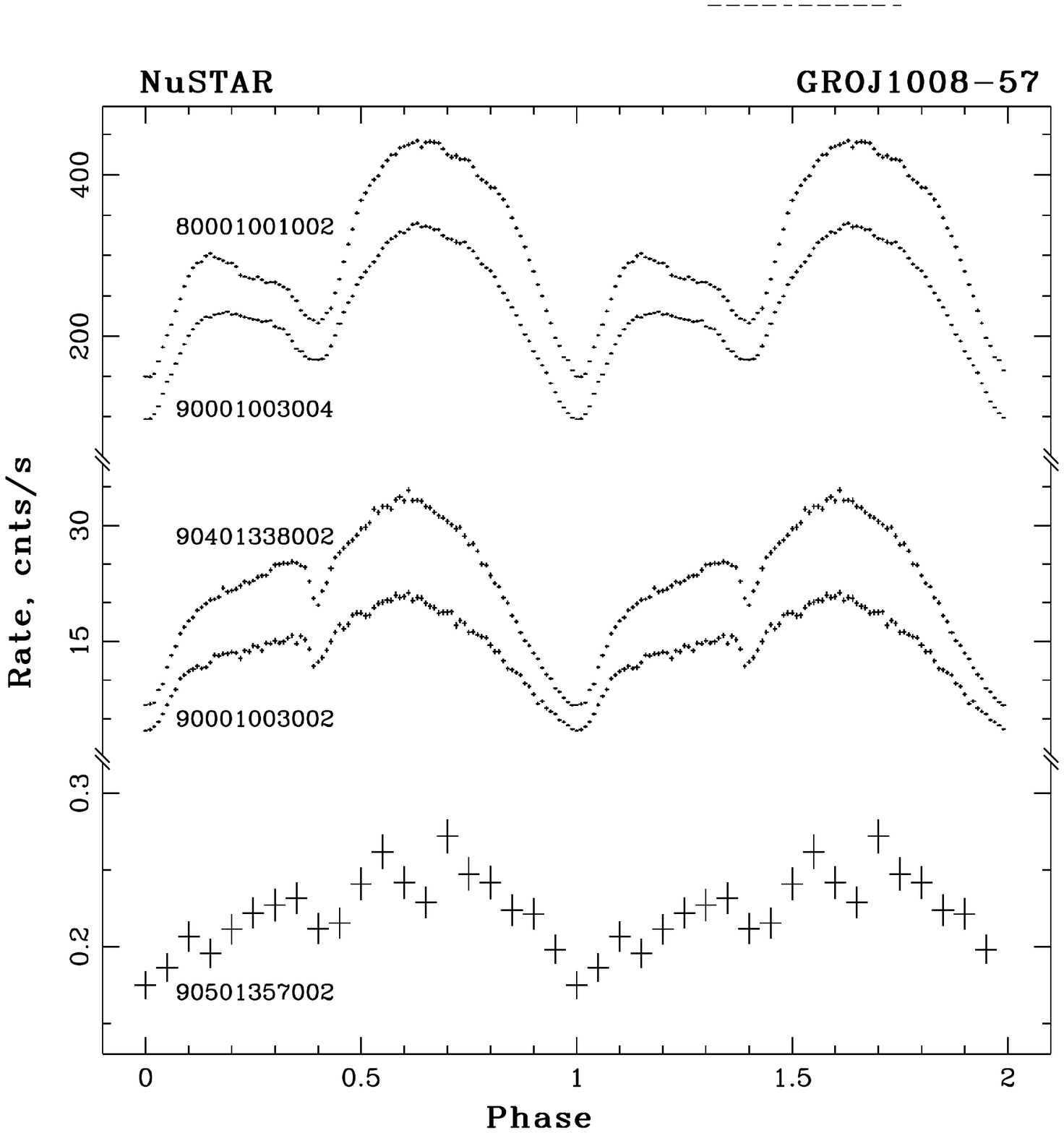}
\caption{Pulse profiles of \gro as seen by \nustar in the energy band of 3-79 keV at different luminosities. }
\label{fig:pprof}
\end{figure}

\section{Results}
\label{sec:res}

The orbital light curve of \gro showing estimated bolometric luminosity over a full orbital cycle is presented in Fig.~\ref{fig:lc} \citep[{\it Swift}/XRT data are adopted from][]{2017A&A...608A..17T}. Observations with \nustar and \art\ are marked by black and light green points, respectively. Note that the observations in Dec, 2019 -- Jan, 2020 have been performed around orbital phases 0.65-0.75, where the source luminosity is minimal at a few times 10$^{34}$~\lum. Below we discuss the temporal and spectral properties of \gro in the low luminosity state and compare them with those observed at higher luminosities.

\subsection{Timing analysis}
\label{subsec:timing}

As was already mentioned in the Introduction, pulsations from \gro in the cold-disk state were not detected using {\it Swift}/XRT and {\it Chandra} data \citep{2017A&A...608A..17T}. One of the main goals of our observations was either to find pulsations or to put a more stringent upper limit on the pulsed fraction which was previously estimated at $\sim20\%$ \citep{2017A&A...608A..17T}.

\begin{figure}
\centering
\vspace{0.6cm}
\includegraphics[width=0.95\columnwidth]{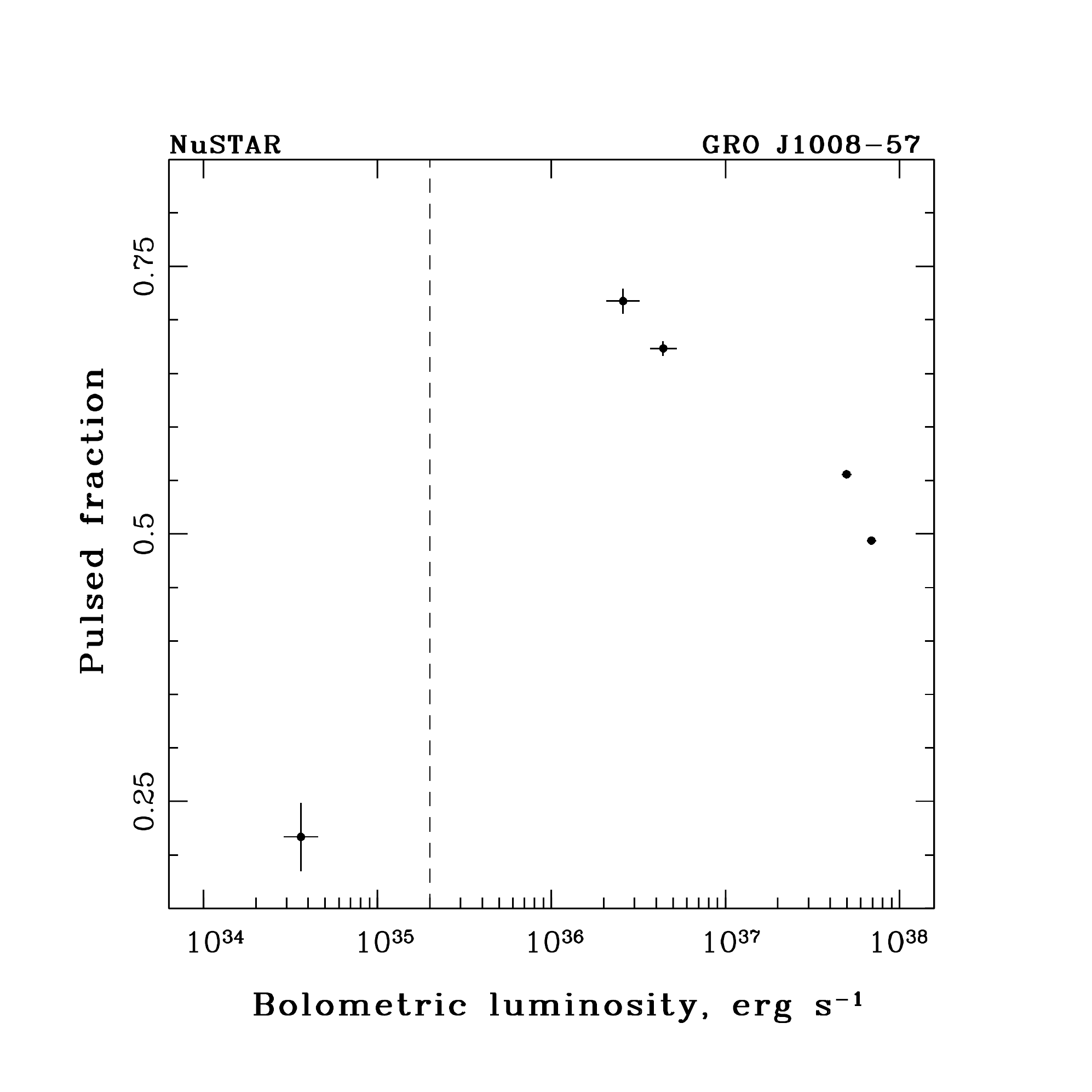}
\caption{Dependence of the pulsed fraction of \gro in the 3-79 keV energy band as a function of luminosity as observed by \nustar. Vertical dashed line represents the luminosity $L_{\rm cold}=2\times10^{35}$~\lum, which corresponds to the switch of the accretion to the cold-disk state \citep{2017A&A...608A..17T}.
}
\label{fig:ppfr}
\end{figure}

For the timing analysis both the barycentric and binary corrections were applied to the light curves using the orbital ephemerides reported by \citet{2013A&A...555A..95K}.
As a result we were able to significantly detect pulsations in the low state with the period of $P\simeq93.2$~s in both \art\ and \nustar\ datasets (see Table~\ref{tab:data} and Fig.\,\ref{fig:artima}, middle panel). To compare obtained values with measured ones in brighter states and to construct corresponding pulse profiles the same analysis was also performed for other four \nustar observations. The results are summarized in Table~\ref{tab:data}. Uncertainties for the pulse period values were determined from the simulated light curves using epoch folding on bootstraped lightcurves as described in detail by \citet{2013AstL...39..375B}.

The pulse profile of \gro in the bright state exhibits two-peaks structure with a relatively high pulsed fraction. The evolution of the pulse profile as a function of luminosity based on the \nustar\ data is presented in Fig.~\ref{fig:pprof}. Some minor variations can be observed, mainly in the first peak. However, in spite of three orders of magnitude difference in the observed source flux, the pulse profile shape remains remarkably stable.

In contrast to that, the pulsed fraction\footnote{defined as $\mathrm{PF}=(F_\mathrm{max}-F_\mathrm{min})/(F_\mathrm{max}+F_\mathrm{min})$, where $F_\mathrm{max}$ and $F_\mathrm{min}$ are maximum and minimum fluxes in the pulse profile, respectively} demonstrates more puzzling behaviour as a function of the source luminosity (Fig.~\ref{fig:ppfr}).
It is seen from the figure, that in the bright state the pulse fraction has relatively high values depending on the flux from $\sim70\%$ to $\sim50\%$. Moreover, at high luminosity levels it also appears to anti-correlate with the flux, which is typical for the majority of XRPs in the hard X-ray band \citep{2009AstL...35..433L}. However, at some luminosity (probably coinciding with the transitional luminosity to the cold-disk accretion, although available data does not allow to determine this time accurately) this dependence breaks down resulting in a very strong drop of the pulsed fraction to $22\pm3$\%. Note, that the pulsed fraction measured by \art\ in the 4-12 keV energy range is approximately the same $20\pm5$\%.

\subsection{Spectral analysis}
\label{subsec:spectra}

Spectra of XRPs provide an important information about the physical mechanisms of the formation of the radiation under extreme conditions in the vicinity of neutron stars and its subsequent interaction with the surrounding matter in strong magnetic fields. These processes are quite complex and not fully understood yet, which makes it difficult to develop physically motivated spectral models.

\begin{figure}
\centering
\includegraphics[width=\columnwidth,bb=35 180 550 710]{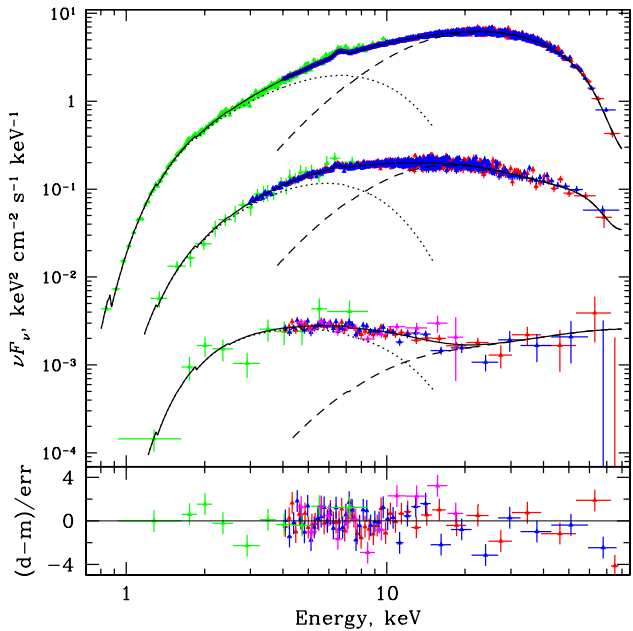}
\caption{{\bf Unfolded} spectra of \gro obtained with \nustar (black and red points), \swiftxrt (green points) and \art\ (magenta points) in different luminosity states. Only three spectra (\nustar\ ObsIDs are 80001001002, 900010003004 and 90501357002) of five are shown for clarity. Corresponding residuals for the lowest luminosity state of \gro\ are shown at the lower panel.}
\label{fig:spectra}
\end{figure}

Therefore, spectra of XRPs have long been described by one-component phenomenological models  (such as a power law with a high energy exponential cutoff, a.k.a. {\sc highcut} or {\sc cutoffpl} in {\sc xspec}), modified by the absorption at lower energies and, in some cases, by the cyclotron absorption lines \citep[see, e.g.,][]{1983ApJ...270..711W,2002ApJ...580..394C,2005AstL...31..729F,2015A&ARv..23....2W}. Expanding the available energy range and increasing the sensitivity of modern instruments made it possible to study spectra of accreting neutron stars in a wide range of luminosities including periods of quiescence. It required the development of new approaches to the analysis of the spectral information and the use of multicomponent models to describe the continuum, although, as before, these models largely remain phenomenological.

The observed spectrum of \gro\ was usually described with the aforementioned phenomenological models, like a cutoff power-law, Fermi–Dirac cutoff powerlaw, or combination of negative and positive powerlaws with an exponential cutoff \citep[see, e.g.,][]{2007MNRAS.378.1427C,2014PASJ...66...59Y} modified by additional components to account for the interstellar absorption, soft excess, fluorescent iron line and cyclotron line around 80\,keV  \citep[see, e.g.,][]{2013A&A...555A..95K, 2014ApJ...792..108B}.

\begin{table}
\small
        \begin{center}
        \caption{Best-fitting results for the \gro broadband spectra in different luminosity states}
        \label{tab:spe}
        \begin{tabular}{lcc}
\hline
\hline
  Parameter$^{a}$  & Low-energy part & High-energy part \\
\hline
\multicolumn{3}{c}{{\sc 80001001002}}\\
\hline
$T_0$, keV             &  $0.30\pm0.05$ &  $3.19\pm0.15$ \\
$T_{\rm p}$, keV       &  $1.89\pm0.05$ &  $8.87\pm0.15$   \\
$\tau_{\rm p}$         &  $12.8\pm0.3$  &  $3.73\pm0.12$ \\
$N_{\rm H}$            &  \multicolumn{2}{c}{$1.14\pm0.10$}  \\
$E_{\rm Fe}$, keV      &  \multicolumn{2}{c}{$6.55\pm0.01$}  \\
$\sigma_{\rm Fe}$, keV &  \multicolumn{2}{c}{$0.28\pm0.01$}  \\
$\tau_{\rm cyc}$,       &  \multicolumn{2}{c}{$1.23\pm0.18$}  \\
$C_{\rm B}$            &  \multicolumn{2}{c}{$1.030\pm0.001$} \\
$C_{\rm XRT}$          &  \multicolumn{2}{c}{$1.132\pm0.005$} \\
$F_{\rm X}$, \flux  &  \multicolumn{2}{c}{$2.08\times10^{-8}$ } \\
$\chi^2$ (d.o.f.)    &  \multicolumn{2}{c}{1.07 (3236)} \\
\hline
\multicolumn{3}{c}{{\sc 90001003004}}\\
\hline
$T_0$, keV             &  $0.28\pm0.06$ &  $4.28\pm0.25$ \\
$T_{\rm p}$, keV       &  $2.22\pm0.06$ &  $12.0\pm1.4$ \\
$\tau_{\rm p}$         &  $11.87\pm0.30$ & $2.24\pm0.41$ \\
$N_{\rm H}$            &  \multicolumn{2}{c}{$1.36\pm0.09$}  \\
$E_{\rm Fe}$, keV      &  \multicolumn{2}{c}{$6.59\pm0.01$}  \\
$\sigma_{\rm Fe}$, keV &  \multicolumn{2}{c}{$0.28\pm0.01$}  \\
$\tau_{\rm cyc}$,       &  \multicolumn{2}{c}{$2.14\pm0.36$}  \\
$C_{\rm B}$            &  \multicolumn{2}{c}{$1.015\pm0.001$} \\
$C_{\rm XRT}$          &  \multicolumn{2}{c}{$1.297\pm0.006$} \\
$F_{\rm X}$,  \flux  &  \multicolumn{2}{c}{$1.50\times10^{-8}$} \\
$\chi^2$ (d.o.f.)    &  \multicolumn{2}{c}{1.10 (2702)} \\
\hline
\multicolumn{3}{c}{{\sc 90401338002}}\\
\hline
$T_0$, keV             &  $0.33\pm0.53$  &  $3.36\pm0.20$ \\
$T_{\rm p}$, keV       &  $1.73\pm0.07$  &  $14.8\pm2.7$ \\
$\tau_{\rm p}$         &  $13.8\pm3.6$  & $1.7\pm0.4$ \\
$N_{\rm H}$            &  \multicolumn{2}{c}{$1.45\pm0.71$}  \\
$E_{\rm Fe}$, keV      &  \multicolumn{2}{c}{$6.34\pm0.02$}  \\
$\sigma_{\rm Fe}$, keV &  \multicolumn{2}{c}{$0.2$ (fix)}  \\
$\tau_{\rm cyc}$,       &  \multicolumn{2}{c}{$1.83\pm0.56$}  \\
$C_{\rm B}$            &  \multicolumn{2}{c}{$1.034\pm0.002$} \\
$F_{\rm X}$,  \flux  &  \multicolumn{2}{c}{$1.34\times10^{-9}$} \\
$\chi^2$ (d.o.f.)    &  \multicolumn{2}{c}{1.01 (1783)} \\
\hline
\multicolumn{3}{c}{{\sc 90001003002}}\\
\hline
$T_0$, keV             &  $0.74\pm0.23$ &  $3.20\pm0.32$ \\
$T_{\rm p}$, keV       &  $1.76\pm0.27$ &  $20\pm16$ \\
$\tau_{\rm p}$         &  $11.8\pm4.1$  & $1.2\pm1.2$ \\
$N_{\rm H}$            &  \multicolumn{2}{c}{$0.66\pm0.29$}  \\
$E_{\rm Fe}$, keV      &  \multicolumn{2}{c}{$6.4$ (fix)}  \\
$\sigma_{\rm Fe}$, keV &  \multicolumn{2}{c}{$0.2$ (fix)}  \\
$\tau_{\rm cyc}$,       &  \multicolumn{2}{c}{$0.8\pm1.0$}  \\
$C_{\rm B}$            &  \multicolumn{2}{c}{$1.027\pm0.004$} \\
$C_{\rm XRT}$          &  \multicolumn{2}{c}{$1.120\pm0.042$} \\
$F_{\rm X}$,  \flux  &  \multicolumn{2}{c}{$8.11\times10^{-10}$} \\
$\chi^2$ (d.o.f.)    &  \multicolumn{2}{c}{1.00 (1356)} \\
\hline
\end{tabular}
\end{center}
\end{table}

\begin{table}
\small
        \begin{center}
        \begin{tabular}{lcc}
\hline
  Parameter$^{a}$  & Low-energy part & High-energy part \\
\hline
\multicolumn{3}{c}{{\sc 90501357002}}\\
\hline
$T_0$, keV             &  $0.25\pm0.53$ &  $>5$ \\
$T_{\rm p}$, keV       &  $3.5\pm3.1$   &  $23\pm19$ \\
$\tau_{\rm p}$         &  $4.4\pm1.6$   & $>5.5$ \\
$N_{\rm H}$            &  \multicolumn{2}{c}{$2.5\pm1.5$}  \\
$C_{\rm B}$            &  \multicolumn{2}{c}{$1.042\pm0.022$} \\
$C_{\rm XRT}$          &  \multicolumn{2}{c}{$0.84\pm0.11$} \\
$C_{\rm ART-XC}$       &  \multicolumn{2}{c}{$0.65\pm0.02$} \\
$F_{\rm X}$,  \flux  &  \multicolumn{2}{c}{$1.50\times10^{-11}$} \\
$\chi^2$ (d.o.f.)    &  \multicolumn{2}{c}{1.01 (1592)} \\
\hline
    \end{tabular}
    \end{center}
	\begin{tablenotes}
        \item $^{a}$ Here $T_{\rm p}$, $\tau_{\rm p}$ and $T_0$ are the plasma temperature, plasma optical depth, temperature of the seed photons for the {\sc comptt} model. Fluxes are given  in the 0.5--100 keV energy range.
        \end{tablenotes}
\end{table}

Recent measurements of the \gro\ broadband spectrum during an outburst with the {\it HMXT} observatory showed that it can be described by two-components (power law with high-energy exponential cutoff plus black body) absorbed continuum model with the cyclotron absorption line at the energy of $\sim90$ keV \citep{2020ApJ...899L..19G}. It is important to note that the line was robustly detected and its parameters were measured with very high significance.

Following previous works, we tried to describe the continuum of the source in a wide range of luminosities with a combination of two components. All of these combinations were modified by the low-energy absorption in the form of {\sc phabs} model in the {\sc xspec} package and by the cyclotron absorption line (in the form of the {\sc gabs} model in the {\sc xspec} package) with the energy and width fixed at values obtained by \citep{2020ApJ...899L..19G}. The line depth was a free parameter. Formally, the line energy is outside of the working energy range of the \nustar observatory, but due to the significant width and depth the absorption feature affects the cutoff energy and thus need to be included in the model.

We fitted the source continuum spectra with different combination
of {\sc xspec} models  previously used by other authors: {\sc bbody+highcut} , {\sc bbody+cutoffpl}, {\sc highcut + gaussian}, {\sc highcut + comptt}, {\sc bbody+comptt}, {\sc comptt+comptt}.  It was found that the combination of the two Comptonization components best describes the source spectrum in all states.
Note, that earlier two Comptonization components were proposed by \cite{2009A&A...498..825F} to fit the spectra of super-critical X-ray pulsar 4U~0115+63 where the model was motivated to account for the influence of bulk and thermal Comptonization in accretion column supported by radiation pressure \citep{2007ApJ...654..435B}. This model, however, is not applicable in the case of sub-critical accretion, which considered in this paper.

To take into account a strong emission feature associated with the fluorescent iron emission line around 6.4 keV, a corresponding component in the Gaussian form was added also to the model in line with the photo-absorption and cyclotron
resonant scattering feature discussed above. The final best-fit parameters (including model flux) are summarized in Table \,\ref{tab:spe}. The energy spectra of \gro are shown in Fig.\,\ref{fig:spectra}.

The source and background spectra from both the
FPMA and FPMB modules of \nustar as well as from the
{\it Swift}/XRT and {\it SRG}/ART-XC telescopes were used for simultaneous fitting. To take into account the uncertainty in the {\it NuSTAR} modules calibrations and potentially slightly different source flux during the (not strictly simultaneous) observations by different observatories, cross-calibration constants between them were included in all spectral models. The $C_{\rm B}$, $C_{\rm XRT}$ and $C_{\rm ART-XC}$ constants from Table\,\ref{tab:spe} correspond to the cross-calibrations of the FPMB module, the XRT telescope to the FPMA module and the ART-XC telescope to the FPMA module, respectively. The value of $C_{\rm ART-XC}$ reflects the source flux change between ART-XC and {\it NuSTAR} observations connected either with its slow brightening towards the next outburst or with its local short-term variability. Nevertheless the spectrum remains practically unchanged that allow us to fit ART-XC and {\it NuSTAR} data simultaneously.

\section{Discussion and conclusion}
\label{sec:conc}

In this work we presented results of the first dedicated broadband observations of \gro in the low state using data of \nustar and \art. Timing analysis from both instruments resulted in the significant detection of pulsations from the source with the pulsed fraction of $\sim20\%$, that is several times lower than in the bright state.

One of the possible explanation of such a behaviour is an increase of the emitting region on the NS surface as a result of the deeper penetration of the cold disk into the pulsar magnetosphere.
More detailed discussion of that requires model of the interaction of the cold disk with magnetic field and is out of scope of the current paper.

Another possibility is related to the structure of a NS atmosphere at low mass accretion rates. According to numerical simulations, the upper layer of the atmosphere is overheated up to temperatures $\sim 100\,{\rm keV}$ \citep{2018A&A...619A.114S,2020arXiv200613596M}.
The photons which leave the atmosphere at larger angles to the normal are originated (or experience the last scattering) from smaller optical depth in the atmosphere.
In the case of temperature increasing with the optical depth, the intensity of radiation tends to be larger along the normal.
In the case of the inverse temperature structure (when the upper layers are hotter than the underlying ones), the intensity increases with the angle to the normal and, thus,
the beam pattern generated by a hot spot with an overheated upper layer is expected to be compressed in the direction perpendicular to the stellar surface.
Such a modification of the beam pattern can result in a reduction of the pulsed fraction.

The qualitative similarity of the pulse profiles (see Fig.\,\ref{fig:pprof}) in different luminosity states allows us to speculate that in all cases, we see the source in the same regime of accretion \citep{1976MNRAS.175..395B,1973A&A....25..233G,2015MNRAS.447.1847M}. Accounting for a strong magnetic field strength at the NS surface, we conclude that the source is still in a sub-critical regime even at $L\sim 10^{38}\,{\rm erg\,s^{-1}}$. It puts valuable limitations to the critical luminosity at extremely strong field strength: $L_{\rm crit}>10^{38}\,{\rm erg\,s^{-1}}$ for $B\simeq 7\times 10^{12}\,{\rm G}$.

We were able for the first time to trace the evolution of the broadband spectrum of the source on a scale of three orders of magnitude in luminosity. It was shown that in all states it can be adequately described by a two-component model with a clear division of the components into soft and hard ones. The spectrum in the high state is in good agreement with the results of previous studies, in which the classical 'power-law with high energy cutoff' model and an additional blackbody component was used. At the lowest state, the spectrum shape becomes a double-hump one. Such a shape is already proved to be typical for XRPs with low mass accretion rate \citep{2012A&A...540L...1D,2019MNRAS.483L.144T,2019MNRAS.487L..30T,VITYA2020}. The tentative peak energy of the hard spectral component is consistent with the position of the cyclotron line at about 80--90 keV, but its temperature is not limited due to low statistics. This is in line with the explanation of this component due to the emission of cyclotron photons in the NS atmosphere caused by collisional excitation of electrons to upper Landau levels and consequent Comptonization by hot electron gas \citep{2020arXiv200613596M}.

\section*{Acknowledgements}

We thanks to the {\it NuSTAR} team for organising
prompt observations. This work was financially supported by the Russian Science Foundation (grant 19-12-00423).




\bibliography{gro1008}

\begin{thebibliography}{}
\expandafter\ifx\csname natexlab\endcsname\relax\def\natexlab#1{#1}\fi
\providecommand{\url}[1]{\href{#1}{#1}}
\providecommand{\dodoi}[1]{doi:~\href{http://doi.org/#1}{\nolinkurl{#1}}}
\providecommand{\doeprint}[1]{\href{http://ascl.net/#1}{\nolinkurl{http://ascl.net/#1}}}
\providecommand{\doarXiv}[1]{\href{https://arxiv.org/abs/#1}{\nolinkurl{https://arxiv.org/abs/#1}}}

\bibitem[{{Bachetti} {et~al.}(2014){Bachetti}, {Harrison}, {Walton},
  {Grefenstette}, {Chakrabarty}, {F{\"u}rst}, {Barret}, {Beloborodov}, {Boggs},
  {Christensen}, {Craig}, {Fabian}, {Hailey}, {Hornschemeier}, {Kaspi},
  {Kulkarni}, {Maccarone}, {Miller}, {Rana}, {Stern}, {Tendulkar}, {Tomsick},
  {Webb}, \& {Zhang}}]{2014Natur.514..202B}
{Bachetti}, M., {Harrison}, F.~A., {Walton}, D.~J., {et~al.} 2014, \nat, 514,
  202, \dodoi{10.1038/nature13791}

\bibitem[{{Basko} \& {Sunyaev}(1976)}]{1976MNRAS.175..395B}
{Basko}, M.~M., \& {Sunyaev}, R.~A. 1976, \mnras, 175, 395,
  \dodoi{10.1093/mnras/175.2.395}

\bibitem[{{Becker} \& {Wolff}(2007)}]{2007ApJ...654..435B}
{Becker}, P.~A., \& {Wolff}, M.~T. 2007, \apj, 654, 435, \dodoi{10.1086/509108}

\bibitem[{{Bellm} {et~al.}(2014){Bellm}, {F{\"u}rst}, {Pottschmidt}, {Tomsick},
  {Boggs}, {Chakrabarty}, {Christensen}, {Craig}, {Hailey}, {Harrison},
  {Stern}, {Walton}, {Wilms}, \& {Zhang}}]{2014ApJ...792..108B}
{Bellm}, E.~C., {F{\"u}rst}, F., {Pottschmidt}, K., {et~al.} 2014, \apj, 792,
  108, \dodoi{10.1088/0004-637X/792/2/108}

\bibitem[{{Boldin} {et~al.}(2013){Boldin}, {Tsygankov}, \&
  {Lutovinov}}]{2013AstL...39..375B}
{Boldin}, P.~A., {Tsygankov}, S.~S., \& {Lutovinov}, A.~A. 2013, Astronomy
  Letters, 39, 375, \dodoi{10.1134/S1063773713060029}

\bibitem[{{Brazier}(1994)}]{1994MNRAS.268..709B}
{Brazier}, K.~T.~S. 1994, \mnras, 268, 709, \dodoi{10.1093/mnras/268.3.709}

\bibitem[{{Campana} {et~al.}(2002){Campana}, {Stella}, {Israel}, {Moretti},
  {Parmar}, \& {Orlandini}}]{2002ApJ...580..389C}
{Campana}, S., {Stella}, L., {Israel}, G.~L., {et~al.} 2002, \apj, 580, 389,
  \dodoi{10.1086/343074}

\bibitem[{{Coburn} {et~al.}(2002){Coburn}, {Heindl}, {Rothschild}, {Gruber},
  {Kreykenbohm}, {Wilms}, {Kretschmar}, \& {Staubert}}]{2002ApJ...580..394C}
{Coburn}, W., {Heindl}, W.~A., {Rothschild}, R.~E., {et~al.} 2002, \apj, 580,
  394, \dodoi{10.1086/343033}

\bibitem[{{Coe} {et~al.}(2007){Coe}, {Bird}, {Hill}, {McBride}, {Schurch},
  {Galache}, {Wilson}, {Finger}, {Buckley}, \&
  {Romero-Colmenero}}]{2007MNRAS.378.1427C}
{Coe}, M.~J., {Bird}, A.~J., {Hill}, A.~B., {et~al.} 2007, \mnras, 378, 1427,
  \dodoi{10.1111/j.1365-2966.2007.11878.x}

\bibitem[{{Doroshenko} {et~al.}(2014){Doroshenko}, {Santangelo}, {Doroshenko},
  {Caballero}, {Tsygankov}, \& {Rothschild}}]{2014A&A...561A..96D}
{Doroshenko}, V., {Santangelo}, A., {Doroshenko}, R., {et~al.} 2014, \aap, 561,
  A96, \dodoi{10.1051/0004-6361/201322472}

\bibitem[{{Doroshenko} {et~al.}(2012){Doroshenko}, {Santangelo}, {Kreykenbohm},
  \& {Doroshenko}}]{2012A&A...540L...1D}
{Doroshenko}, V., {Santangelo}, A., {Kreykenbohm}, I., \& {Doroshenko}, R.
  2012, \aap, 540, L1, \dodoi{10.1051/0004-6361/201218878}

\bibitem[{{Doroshenko} {et~al.}(2021){Doroshenko}, {Santangelo}, {Tsygankov},
  \& {Long}}]{VITYA2020}
{Doroshenko}, V., {Santangelo}, A., {Tsygankov}, S.~S., \& {Long}, J. 2021,
  \aap, submitted

\bibitem[{{Doroshenko} {et~al.}(2020){Doroshenko}, {Zhang}, {Santangelo}, {Ji},
  {Tsygankov}, {Mushtukov}, {Qu}, {Zhang}, {Ge}, {Chen}, {Bu}, {Cao}, {Chang},
  {Chen}, {Chen}, {Chen}, {Chen}, {Chen}, {Cui}, {Cui}, {Deng}, {Dong}, {Du},
  {Fu}, {Gao}, {Gao}, {Gao}, {Gu}, {Guan}, {Guo}, {Han}, {Hu}, {Huang}, {Huo},
  {Jia}, {Jiang}, {Jiang}, {Jin}, {Jin}, {Kong}, {Li}, {Li}, {Li}, {Li}, {Li},
  {Li}, {Li}, {Li}, {Li}, {Li}, {Li}, {Li}, {Liang}, {Liao}, {Liu}, {Liu},
  {Liu}, {Liu}, {Liu}, {Liu}, {Liu}, {Lu}, {Lu}, {Lu}, {Luo}, {Ma}, {Meng},
  {Nang}, {Nie}, {Ou}, {Sai}, {Shang}, {Song}, {Song}, {Sun}, {Tan}, {Tao},
  {Tuo}, {Wang}, {Wang}, {Wang}, {Wang}, {Wen}, {Wu}, {Wu}, {Xiao}, {Xiong},
  {Xu}, {Xu}, {Yang}, {Yang}, {Yang}, {Yang}, {Zhang}, {Zhang}, {Zhang},
  {Zhang}, {Zhang}, {Zhang}, {Zhang}, {Zhang}, {Zhang}, {Zhang}, {Zhang},
  {Zhang}, {Zhang}, {Zhang}, {Zhang}, {Zhang}, {Zhang}, {Zhao}, {Zhao}, {Zhao},
  {Zheng}, {Zhu}, {Zhu}, {Zou}, \& {Zhang}}]{2020MNRAS.491.1857D}
{Doroshenko}, V., {Zhang}, S.~N., {Santangelo}, A., {et~al.} 2020, \mnras, 491,
  1857, \dodoi{10.1093/mnras/stz2879}

\bibitem[{{Evans} {et~al.}(2009){Evans}, {Beardmore}, {Page}, {Osborne},
  {O'Brien}, {Willingale}, {Starling}, {Burrows}, {Godet}, {Vetere}, {Racusin},
  {Goad}, {Wiersema}, {Angelini}, {Capalbi}, {Chincarini}, {Gehrels}, {Kennea},
  {Margutti}, {Morris}, {Mountford}, {Pagani}, {Perri}, {Romano}, \&
  {Tanvir}}]{2009MNRAS.397.1177E}
{Evans}, P.~A., {Beardmore}, A.~P., {Page}, K.~L., {et~al.} 2009, \mnras, 397,
  1177, \dodoi{10.1111/j.1365-2966.2009.14913.x}

\bibitem[{{Ferrigno} {et~al.}(2009){Ferrigno}, {Becker}, {Segreto}, {Mineo}, \&
  {Santangelo}}]{2009A&A...498..825F}
{Ferrigno}, C., {Becker}, P.~A., {Segreto}, A., {Mineo}, T., \& {Santangelo},
  A. 2009, \aap, 498, 825, \dodoi{10.1051/0004-6361/200809373}

\bibitem[{{Filippova} {et~al.}(2005){Filippova}, {Tsygankov}, {Lutovinov}, \&
  {Sunyaev}}]{2005AstL...31..729F}
{Filippova}, E.~V., {Tsygankov}, S.~S., {Lutovinov}, A.~A., \& {Sunyaev}, R.~A.
  2005, Astronomy Letters, 31, 729, \dodoi{10.1134/1.2123288}

\bibitem[{{F{\"u}rst} {et~al.}(2016){F{\"u}rst}, {Walton}, {Harrison}, {Stern},
  {Barret}, {Brightman}, {Fabian}, {Grefenstette}, {Madsen}, {Middleton},
  {Miller}, {Pottschmidt}, {Ptak}, {Rana}, \& {Webb}}]{2016ApJ...831L..14F}
{F{\"u}rst}, F., {Walton}, D.~J., {Harrison}, F.~A., {et~al.} 2016, \apjl, 831,
  L14, \dodoi{10.3847/2041-8205/831/2/L14}

\bibitem[{{Ge} {et~al.}(2020){Ge}, {Ji}, {Zhang}, {Santangelo}, {Liu},
  {Doroshenko}, {Staubert}, {Qu}, {Zhang}, {Lu}, {Song}, {Li}, {Tao}, {Xu},
  {Cao}, {Chen}, {Bu}, {Cai}, {Chang}, {Chen}, {Chen}, {Chen}, {Chen}, {Chen},
  {Cui}, {Cui}, {Deng}, {Dong}, {Du}, {Fu}, {Gao}, {Gao}, {Gao}, {Gu}, {Guan},
  {Guo}, {Han}, {Huang}, {Huo}, {Jia}, {Jiang}, {Jiang}, {Jin}, {Jin}, {Kong},
  {Li}, {Li}, {Li}, {Li}, {Li}, {Li}, {Li}, {Li}, {Li}, {Li}, {Liang}, {Liao},
  {Liu}, {Liu}, {Liu}, {Liu}, {Liu}, {Lu}, {Lu}, {Luo}, {Luo}, {Ma}, {Meng},
  {Nang}, {Nie}, {Ou}, {Sai}, {Shang}, {Song}, {Sun}, {Tan}, {Tuo}, {Wang},
  {Wang}, {Wang}, {Wang}, {Wang}, {Wang}, {Wang}, {Wen}, {Wu}, {Wu}, {Wu},
  {Xiao}, {Xiao}, {Xiong}, {Xu}, {Yang}, {Yang}, {Yang}, {Yang}, {Yi}, {Yin},
  {You}, {Zhang}, {Zhang}, {Zhang}, {Zhang}, {Zhang}, {Zhang}, {Zhang},
  {Zhang}, {Zhang}, {Zhang}, {Zhang}, {Zhang}, {Zhang}, {Zhang}, {Zhang},
  {Zhang}, {Zhao}, {Zhao}, {Zheng}, {Zheng}, {Zhou}, {Zhou}, {Zhuang}, {Zhu},
  \& {Zhu}}]{2020ApJ...899L..19G}
{Ge}, M.~Y., {Ji}, L., {Zhang}, S.~N., {et~al.} 2020, \apjl, 899, L19,
  \dodoi{10.3847/2041-8213/abac05}

\bibitem[{{Gnedin} \& {Sunyaev}(1973)}]{1973A&A....25..233G}
{Gnedin}, Y.~N., \& {Sunyaev}, R.~A. 1973, \aap, 25, 233

\bibitem[{{Harrison} {et~al.}(2013){Harrison}, {Craig}, {Christensen},
  {Hailey}, {Zhang}, {Boggs}, {Stern}, {Cook}, {Forster}, {Giommi},
  {Grefenstette}, {Kim}, {Kitaguchi}, {Koglin}, {Madsen}, {Mao}, {Miyasaka},
  {Mori}, {Perri}, {Pivovaroff}, {Puccetti}, {Rana}, {Westergaard}, {Willis},
  {Zoglauer}, {An}, {Bachetti}, {Barri{\`e}re}, {Bellm}, {Bhalerao},
  {Brejnholt}, {Fuerst}, {Liebe}, {Markwardt}, {Nynka}, {Vogel}, {Walton},
  {Wik}, {Alexander}, {Cominsky}, {Hornschemeier}, {Hornstrup}, {Kaspi},
  {Madejski}, {Matt}, {Molendi}, {Smith}, {Tomsick}, {Ajello}, {Ballantyne},
  {Balokovi{\'c}}, {Barret}, {Bauer}, {Blandford}, {Brandt}, {Brenneman},
  {Chiang}, {Chakrabarty}, {Chenevez}, {Comastri}, {Dufour}, {Elvis}, {Fabian},
  {Farrah}, {Fryer}, {Gotthelf}, {Grindlay}, {Helfand}, {Krivonos}, {Meier},
  {Miller}, {Natalucci}, {Ogle}, {Ofek}, {Ptak}, {Reynolds}, {Rigby},
  {Tagliaferri}, {Thorsett}, {Treister}, \& {Urry}}]{2013ApJ...770..103H}
{Harrison}, F.~A., {Craig}, W.~W., {Christensen}, F.~E., {et~al.} 2013, \apj,
  770, 103, \dodoi{10.1088/0004-637X/770/2/103}

\bibitem[{{Israel} {et~al.}(2017){Israel}, {Belfiore}, {Stella}, {Esposito},
  {Casella}, {De Luca}, {Marelli}, {Papitto}, {Perri}, {Puccetti}, {Castillo},
  {Salvetti}, {Tiengo}, {Zampieri}, {D'Agostino}, {Greiner}, {Haberl},
  {Novara}, {Salvaterra}, {Turolla}, {Watson}, {Wilms}, \&
  {Wolter}}]{2017Sci...355..817I}
{Israel}, G.~L., {Belfiore}, A., {Stella}, L., {et~al.} 2017, Science, 355,
  817, \dodoi{10.1126/science.aai8635}

\bibitem[{{Kuehnel} {et~al.}(2012){Kuehnel}, {Mueller}, {Kreykenbohm}, {Wilms},
  {Pottschmidt}, {Fuerst}, {Rothschild}, {Caballero}, {Klochkov}, {Staubert},
  {Suchy}, {Kretschmar}, {Ferrigno}, {Torrejon}, \&
  {Martinez-Nunez}}]{2012ATel.4564....1K}
{Kuehnel}, M., {Mueller}, S., {Kreykenbohm}, I., {et~al.} 2012, The
  Astronomer's Telegram, 4564, 1

\bibitem[{{K{\"u}hnel} {et~al.}(2013){K{\"u}hnel}, {M{\"u}ller}, {Kreykenbohm},
  {F{\"u}rst}, {Pottschmidt}, {Rothschild}, {Caballero}, {Grinberg},
  {Sch{\"o}nherr}, {Shrader}, {Klochkov}, {Staubert}, {Ferrigno},
  {Torrej{\'o}n}, {Mart{\'\i}nez-N{\'u}{\~n}ez}, \&
  {Wilms}}]{2013A&A...555A..95K}
{K{\"u}hnel}, M., {M{\"u}ller}, S., {Kreykenbohm}, I., {et~al.} 2013, \aap,
  555, A95, \dodoi{10.1051/0004-6361/201321203}

\bibitem[{{Levine} \& {Corbet}(2006)}]{2006ATel..940....1L}
{Levine}, A.~M., \& {Corbet}, R. 2006, The Astronomer's Telegram, 940, 1

\bibitem[{{Lutovinov} \& {Tsygankov}(2009)}]{2009AstL...35..433L}
{Lutovinov}, A.~A., \& {Tsygankov}, S.~S. 2009, Astronomy Letters, 35, 433,
  \dodoi{10.1134/S1063773709070019}

\bibitem[{{Lutovinov} {et~al.}(2019){Lutovinov}, {Tsygankov}, {Karasev},
  {Molkov}, \& {Doroshenko}}]{2019MNRAS.485..770L}
{Lutovinov}, A.~A., {Tsygankov}, S.~S., {Karasev}, D.~I., {Molkov}, S.~V., \&
  {Doroshenko}, V. 2019, \mnras, 485, 770, \dodoi{10.1093/mnras/stz437}

\bibitem[{{Lutovinov} {et~al.}(2017){Lutovinov}, {Tsygankov}, {Krivonos},
  {Molkov}, \& {Poutanen}}]{2017ApJ...834..209L}
{Lutovinov}, A.~A., {Tsygankov}, S.~S., {Krivonos}, R.~A., {Molkov}, S.~V., \&
  {Poutanen}, J. 2017, \apj, 834, 209, \dodoi{10.3847/1538-4357/834/2/209}

\bibitem[{{Mushtukov} {et~al.}(2020){Mushtukov}, {Suleimanov}, {Tsygankov}, \&
  {Portegies Zwart}}]{2020arXiv200613596M}
{Mushtukov}, A.~A., {Suleimanov}, V.~F., {Tsygankov}, S.~S., \& {Portegies
  Zwart}, S. 2020, arXiv e-prints, arXiv:2006.13596.
\newblock \doarXiv{2006.13596}

\bibitem[{{Mushtukov} {et~al.}(2015){Mushtukov}, {Suleimanov}, {Tsygankov}, \&
  {Poutanen}}]{2015MNRAS.447.1847M}
{Mushtukov}, A.~A., {Suleimanov}, V.~F., {Tsygankov}, S.~S., \& {Poutanen}, J.
  2015, \mnras, 447, 1847, \dodoi{10.1093/mnras/stu2484}

\bibitem[{{Pavlinsky} {et~al.}(2021){Pavlinsky}, {Tkachenko}, {Levin}, \& {et
  al.}}]{ART2021}
{Pavlinsky}, M., {Tkachenko}, A., {Levin}, V., \& {et al.} 2021, \aap,
  submitted

\bibitem[{{Riquelme} {et~al.}(2012){Riquelme}, {Torrej{\'o}n}, \&
  {Negueruela}}]{2012A&A...539A.114R}
{Riquelme}, M.~S., {Torrej{\'o}n}, J.~M., \& {Negueruela}, I. 2012, \aap, 539,
  A114, \dodoi{10.1051/0004-6361/201117738}

\bibitem[{{Rodr{\'\i}guez Castillo} {et~al.}(2020){Rodr{\'\i}guez Castillo},
  {Israel}, {Belfiore}, {Bernardini}, {Esposito}, {Pintore}, {De Luca},
  {Papitto}, {Stella}, {Tiengo}, {Zampieri}, {Bachetti}, {Brightman},
  {Casella}, {D'Agostino}, {Dall'Osso}, {Earnshaw}, {F{\"u}rst}, {Haberl},
  {Harrison}, {Mapelli}, {Marelli}, {Middleton}, {Pinto}, {Roberts},
  {Salvaterra}, {Turolla}, {Walton}, \& {Wolter}}]{2020ApJ...895...60R}
{Rodr{\'\i}guez Castillo}, G.~A., {Israel}, G.~L., {Belfiore}, A., {et~al.}
  2020, \apj, 895, 60, \dodoi{10.3847/1538-4357/ab8a44}

\bibitem[{{Semena} {et~al.}(2019){Semena}, {Lutovinov}, {Mereminskiy},
  {Tsygankov}, {Shtykovsky}, {Molkov}, \& {Poutanen}}]{2019MNRAS.490.3355S}
{Semena}, A.~N., {Lutovinov}, A.~A., {Mereminskiy}, I.~A., {et~al.} 2019,
  \mnras, 490, 3355, \dodoi{10.1093/mnras/stz2722}

\bibitem[{{Shrader} {et~al.}(1999){Shrader}, {Sutaria}, {Singh}, \&
  {Macomb}}]{1999ApJ...512..920S}
{Shrader}, C.~R., {Sutaria}, F.~K., {Singh}, K.~P., \& {Macomb}, D.~J. 1999,
  \apj, 512, 920, \dodoi{10.1086/306785}

\bibitem[{{Stollberg} {et~al.}(1993){Stollberg}, {Finger}, {Wilson}, {Harmon},
  {Rubin}, {Zhang}, \& {Fishman}}]{1993IAUC.5836....1S}
{Stollberg}, M.~T., {Finger}, M.~H., {Wilson}, R.~B., {et~al.} 1993, \iaucirc,
  5836, 1

\bibitem[{{Suleimanov} {et~al.}(2018){Suleimanov}, {Poutanen}, \&
  {Werner}}]{2018A&A...619A.114S}
{Suleimanov}, V.~F., {Poutanen}, J., \& {Werner}, K. 2018, \aap, 619, A114,
  \dodoi{10.1051/0004-6361/201833581}

\bibitem[{{Tsygankov} {et~al.}(2019{\natexlab{a}}){Tsygankov}, {Doroshenko},
  {Mushtukov}, {Suleimanov}, {Lutovinov}, \& {Poutanen}}]{2019MNRAS.487L..30T}
{Tsygankov}, S.~S., {Doroshenko}, V., {Mushtukov}, A.~A., {et~al.}
  2019{\natexlab{a}}, \mnras, 487, L30, \dodoi{10.1093/mnrasl/slz079}

\bibitem[{{Tsygankov} {et~al.}(2016){Tsygankov}, {Lutovinov}, {Doroshenko},
  {Mushtukov}, {Suleimanov}, \& {Poutanen}}]{2016A&A...593A..16T}
{Tsygankov}, S.~S., {Lutovinov}, A.~A., {Doroshenko}, V., {et~al.} 2016, \aap,
  593, A16, \dodoi{10.1051/0004-6361/201628236}

\bibitem[{{Tsygankov} {et~al.}(2017{\natexlab{a}}){Tsygankov}, {Mushtukov},
  {Suleimanov}, {Doroshenko}, {Abolmasov}, {Lutovinov}, \&
  {Poutanen}}]{2017A&A...608A..17T}
{Tsygankov}, S.~S., {Mushtukov}, A.~A., {Suleimanov}, V.~F., {et~al.}
  2017{\natexlab{a}}, \aap, 608, A17, \dodoi{10.1051/0004-6361/201630248}

\bibitem[{{Tsygankov} {et~al.}(2019{\natexlab{b}}){Tsygankov}, {Rouco
  Escorial}, {Suleimanov}, {Mushtukov}, {Doroshenko}, {Lutovinov}, {Wijnand s},
  \& {Poutanen}}]{2019MNRAS.483L.144T}
{Tsygankov}, S.~S., {Rouco Escorial}, A., {Suleimanov}, V.~F., {et~al.}
  2019{\natexlab{b}}, \mnras, 483, L144, \dodoi{10.1093/mnrasl/sly236}

\bibitem[{{Tsygankov} {et~al.}(2017{\natexlab{b}}){Tsygankov}, {Wijnands},
  {Lutovinov}, {Degenaar}, \& {Poutanen}}]{2017MNRAS.470..126T}
{Tsygankov}, S.~S., {Wijnands}, R., {Lutovinov}, A.~A., {Degenaar}, N., \&
  {Poutanen}, J. 2017{\natexlab{b}}, \mnras, 470, 126,
  \dodoi{10.1093/mnras/stx1255}

\bibitem[{{Wachter} {et~al.}(1979){Wachter}, {Leach}, \&
  {Kellogg}}]{1979ApJ...230..274W}
{Wachter}, K., {Leach}, R., \& {Kellogg}, E. 1979, \apj, 230, 274,
  \dodoi{10.1086/157084}

\bibitem[{{Walter} {et~al.}(2015){Walter}, {Lutovinov}, {Bozzo}, \&
  {Tsygankov}}]{2015A&ARv..23....2W}
{Walter}, R., {Lutovinov}, A.~A., {Bozzo}, E., \& {Tsygankov}, S.~S. 2015,
  \aapr, 23, 2, \dodoi{10.1007/s00159-015-0082-6}

\bibitem[{{White} {et~al.}(1983){White}, {Swank}, \&
  {Holt}}]{1983ApJ...270..711W}
{White}, N.~E., {Swank}, J.~H., \& {Holt}, S.~S. 1983, \apj, 270, 711,
  \dodoi{10.1086/161162}

\bibitem[{{Wijnands} \& {Degenaar}(2016)}]{2016MNRAS.463L..46W}
{Wijnands}, R., \& {Degenaar}, N. 2016, \mnras, 463, L46,
  \dodoi{10.1093/mnrasl/slw096}

\bibitem[{{Yamamoto} {et~al.}(2014){Yamamoto}, {Mihara}, {Sugizaki},
  {Nakajima}, {Makishima}, \& {Sasano}}]{2014PASJ...66...59Y}
{Yamamoto}, T., {Mihara}, T., {Sugizaki}, M., {et~al.} 2014, \pasj, 66, 59,
  \dodoi{10.1093/pasj/psu028}

\bibitem[{{Yamamoto} {et~al.}(2013){Yamamoto}, {Mihara}, {Sugizaki}, {Sasano},
  {Makishima}, \& {Nakajima}}]{2013ATel.4759....1Y}
---. 2013, The Astronomer's Telegram, 4759, 1

\end{thebibliography}
\bibliographystyle{aasjournal}

\label{lastpage}
\end{document}